\documentclass[aps,twocolumn,showlabels,showrefs,amsmath,amssymb,pre,superscriptaddress, floatfix, colors]{revtex4-1}

\usepackage{amssymb,amsmath,amsfonts}
\usepackage{graphicx,color}
\usepackage[english]{babel}
\usepackage{tabularx}

\usepackage{soul}

\begin{document} 

\title{Influence of Sensorial Delay on Clustering and Swarming}

\author{Rafal Piwowarczyk}
\affiliation{Department of Physics, University of Gothenburg, SE-41296 Gothenburg, Sweden, EU}
\affiliation{Department of Physics, Chalmers University of Technology, SE-41296 Gothenburg, Sweden, EU}

\author{Martin Selin}
\affiliation{Department of Physics, University of Gothenburg, SE-41296 Gothenburg, Sweden, EU}
\affiliation{Department of Physics, Chalmers University of Technology, SE-41296 Gothenburg, Sweden, EU}

\author{Thomas Ihle}
\affiliation{Institute of Physics, University of Greifswald, DE-17489 Greifswald, Germany, EU}

\author{Giovanni Volpe}
\email{giovanni.volpe@physics.gu.se}
\affiliation{Department of Physics, University of Gothenburg, SE-41296 Gothenburg, Sweden, EU}

\date{\today}

\begin{abstract}
We show that sensorial delay alters the collective motion of self-propelling agents with aligning interactions:
In a two-dimensional Vicsek model, short delays enhance the emergence of clusters and swarms, while long or negative delays prevent their formation.
In order to quantify this phenomenon, we introduce a global clustering parameter based on the Voronoi tessellation, which permits us to efficiently measure the formation of clusters.
Thanks to its simplicity, sensorial delay might already play a role in the organization of living organisms and can provide a powerful tool to engineer and dynamically tune the behavior of large ensembles of autonomous robots.
\end{abstract}

\maketitle

Collective motion and pattern formation in systems of self-propelled agents are fascinating phenomena, which have attracted much attention \cite{bechinger2016active,vicsek2012collective}. Systems of interest include, for example, animal flocks \cite{vicsek2012collective,toner2005hydrodynamics,li2008minimal,buhl2006disorder}, chemically powered nanorods \cite{xu2015reversible}, actin networks driven by molecular motors \cite{sanchez2012spontaneous}, robotic swarms \cite{brambilla2013swarm}, and human crowds \cite{moussaid2010walking}. 
Despite occurring on different scales, there are several emergent behaviors that are robust and universal, being in particular independent of the agents constituting the swarm \cite{vicsek2012collective,toner2005hydrodynamics,li2008minimal,buhl2006disorder,xu2015reversible,sanchez2012spontaneous,brambilla2013swarm,moussaid2010walking,nilsson2017metastable}. In the past decades, it has become a challenge for theoretical physics to find minimal dynamical models that capture these features.
In 1987, Reynolds introduced the boids model to simulate the swarm behavior of animals at the macroscale within computer graphics applications \cite{reynolds1987flocks}. Later, in 1995, Vicsek et al. \cite{vicsek1995novel} introduced the Vicsek model, which was the first to consider collective motion in terms of a noise-induced phase transition and, together with its multiple variants, has become one of the most often employed models \cite{czirok2000collective,chate2008modeling,chate2008collective,vicsek2012collective}.
The dynamics of these systems can be influenced by the presence of delay \cite{mijalkov2016engineering,gerlee2017impact,strombom2017anticipation} or multiplicative noise \cite{pesce2013stratonovich,volpe2016effective}.
Using robots, we have recently demonstrated that it is possible to use the delay between the time when an agent senses a signal and the time when it reacts to it (\emph{sensorial delay}) as a new parameter to engineer their large-scale organization \cite{mijalkov2016engineering}. 
 
Here, we show that the presence of a sensorial delay alters the collective motion emerging in the Vicsek model. Considering a collection of self-propelling agents moving with constant speed and tending to align with the average direction of motion of the agents in their local neighbourhood, we show that a short sensorial delay enhances the formation and stability of swarms, while longer or negative delays prevent swarm formation.
We analyze these behaviors using the \emph{global swarming coefficient}, which is standard since the introduction of the Vicsek model \cite{vicsek1995novel}, and by introducing a \emph{global clustering coefficient} based on the Voronoi tessellation, which better captures the formation of swarming clusters.

\begin{figure*}
\includegraphics[width=1.5\columnwidth]{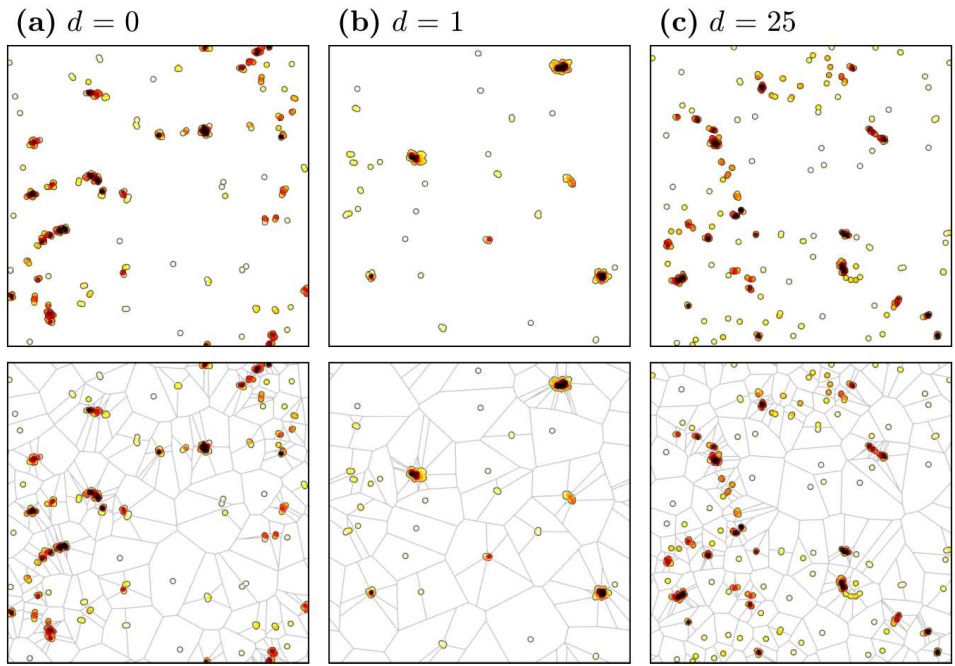}
\caption{
The presence of sensorial delay $d$ influences the clustering and swarming of $N=400$ self-propelling agents (circles, speed $\nu = 3$, detection radius $R=20$) in a 2D Vicsek model (arena size $2\,000 \times 2\,000$ with periodic boundary conditions): 
(a) $d=0$ (standard Vicsek model); (b) $d=1$ enhances the formation of swarms; (c) $d=25$ prevents the emergence of swarms.
The snapshots are taken once the system has reached the steady state (timestep $8\,000$).
In the top row only the agents are shown, while in the bottom row also the borders of the corresponding Voronoi tessellation are shown. The color of the agents denotes the area of the corresponding Voronoi cell going from black (smallest area) to white (largest area).
See also supplementary video 1.
}
\label{fig1}
\end{figure*}

\begin{figure}
\includegraphics[width=1.0\columnwidth]{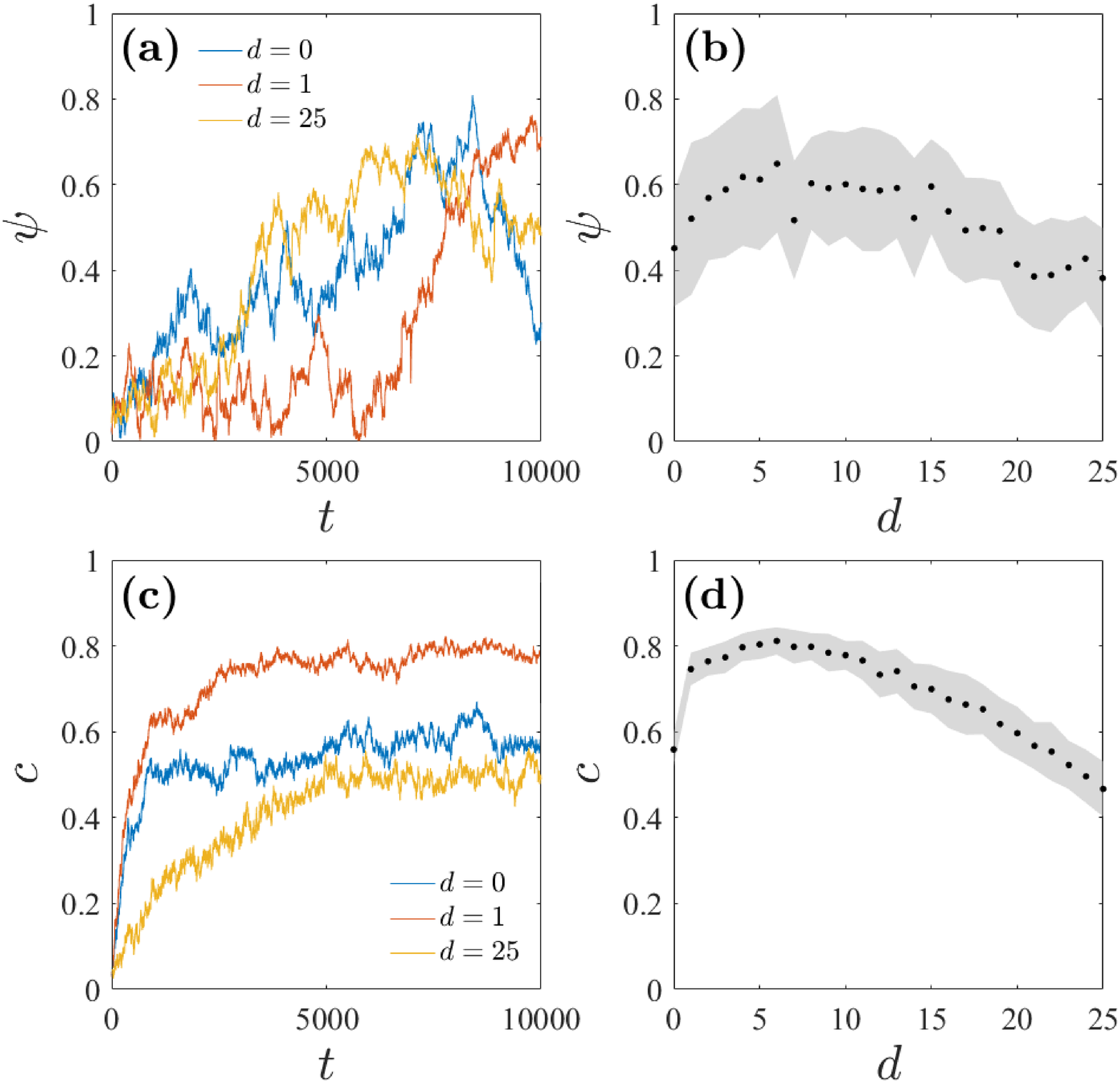}
\caption{
Global alignment and clustering parameters.
(a) The evolution of the global alignment parameter $\psi$ as a function of time does not show a clear difference between delays $d=0$, $1$, and $25$, and 
(b) the average steady-state (i.e., measured after $5\,000$ timesteps) $\psi$ does not show statistically significant differences as a function of $d$ (the shaded area represents a standard deviation over 10 independent runs).
(c) The evolution of the global clustering parameter $c$ shows marked differences between delays $d=0$, $1$, and $25$, and 
(d) the average steady-state (i.e., measured after $5\,000$ timesteps) $c$ clearly depends on $d$ (the shaded area represents a standard deviation over 10 independent runs).
In all cases, we consider $N = 400$ agents with $\nu=3$ and $R=20$ in a square $2\,000 \times 2\,000$ arena with periodic boundary conditions.
}
\label{fig2}
\end{figure}

We consider the standard two-dimensional (2D) Vicsek model \cite{vicsek1995novel}. The self-propelling agents are point-like particles with position ${\bf r}_i$ and orientation $\theta_i$, where $i = 1, ..., N$ and $N$ is the total number of agents. These agents move in a square arena ($2\,000 \times 2\,000$, with periodic boundary conditions) with constant speed $\nu$ and orientation determined by thermal noise and short-range aligning interactions. 
At each timestep, the $i$-th agent moves by $\nu$ ($\nu = 1$, $3$, or $7$ depending on the simulation) along the direction defined by $\theta_i$; the value of $\theta_i$ is set equal to the average direction of the agents within the detection radius $R = 20$ (including the $i$-th agent itself); and, finally, $\theta_i$ is incremented by  a random number drawn from a uniform distribution in $[\frac{-\eta}{2},\frac{\eta}{2}]$, where $\eta = 0.4$ \footnote{The value $\eta = 0.4$ is chosen because it is large enough to prevent formation of permanent clusters, and is also small enough so that the agent behavior is mainly controlled by the aligning interactions.}.
Each simulation run starts with random ${\bf r}_i$ and $\theta_i$, is carried out for $10\,000$ timesteps, and reaches the steady state at most after $5\,000$ timesteps (and typically much sooner).
A crucial parameter in determining the behavior of the system is the density of agents. We consider a system with density of $\delta = 0.0001$ agents per unit area, so that the partner number (i.e., the average number of agents within a detection radius) is $M = \delta \pi R^2 = 0.13$. We have chosen this value in order to achieve a relatively fast equilibration period ($\ll 5\,000$ timesteps) while preventing the appearance of density waves, which occur at significantly higher densities \cite{nagy2007new,bertin2006boltzmann,ihle2011kinetic,ihle2013invasion}. We have verified that density waves occur in our system only when the density is at least ten times larger than that we employed. 

When multiple agents come within a detection radius from each other, they form a local cluster traveling approximately in the same direction and, as time passes, more and more agents are recruited to the cluster, while some agents (usually one by one) leave the cluster, leading to a stationary distribution of cluster sizes.
When the thermal orientational noise is small enough for the aligning interactions to take hold, over time all agents tend to align and travel in the same direction. Importantly, this global alignment occurs despite each agent sensing only its immediate surroundings and having no knowledge of an overall plan. This is traditionally measured by using a \emph{global alignment parameter} given by the modulus of the average normalized velocity \cite{vicsek1995novel}:
\begin{equation} \label{eq:1}
    \psi = {1 \over N} \left| \sum^{N}_{i=1} { {\bf v}_i \over \nu} \right| \; ,
\end{equation}
where ${\bf v}_i$ is the velocity of the $i$-th agent. When $\theta_i$ are completely randomized, $\psi$ is close to zero, while $\psi \approx 1$ when most agents are aligned. A snapshot of the standard Vicsek model is shown in Fig.~\ref{fig1}a (see also supplementary video 1). The corresponding evolution of $\psi$ as a function of the timestep is shown by the blue line in Fig.~\ref{fig2}a: It can be seen that, for the parameters of our model (which are chosen so that the orientational noise is below the threshold necessary for global alignment \cite{vicsek1995novel,vicsek2012collective}), a high degree of alignment is achieved, even though fluctuations persist. 

While a defining ingredient of the Vicsek model is the \emph{instantaneous} alignment of the agents, in realistic systems and applications \emph{delays} are often present due, for example, to the time it takes to acquire, transmit and process sensorial data about the environment surrounding an agent. Therefore, in the following we explore what happens to the Vicsek model when we introduce a delay $d$ between the moment when an agent measures the average orientation within a detection radius and the moment when it changes its own orientation accordingly.
As can be seen in Fig.~\ref{fig1}b (see also supplementary video 1), the introduction of a one-timestep delay ($d=1$) enhances the formation of clusters and the emergence of swarming; in fact, the resulting swarms are more stable than in the case of the standard Vicsek model ($d=0$, Fig.~\ref{fig1}a). 
This can be understood taking into account that, by introducing a small delay, the reorientation of the agents acquires an enhanced persistence time and, therefore, becomes more robust against the orientational noise. 
Interestingly, we have found that a theoretical model taking into account first-order correlations between consecutive timesteps is not affected by the introduction of a delay, so that we conclude that the effects we observed numerically for nonzero delay can only be properly described by including higher-order correlations, which will require the development of non-trivial theoretical tools \footnote{We developed a kinetic theory for a delay of one time step, $d=1$, which is based on the two-time phase-space density of the $N$-agent system. This theory combines the phase-space approach for the regular Vicsek model \cite{ihle2014towards} with Rostoker's kinetic theory from plasma physics \cite{rostoker1961fluctuations}. Within a generalized molecular chaos closure that includes correlations between states at the two different times but neglects equal-time correlations, the phase diagram for spatially homogeneous states was calculated and found to be identical to that of the regular Visek-model with zero delay \cite{ihle2013invasion}. Thus, we expect that the effects we observed numerically for nonzero delay can only be properly described by including correlations in a more realistic way, which will be left for future work.}.

Increasing the value of $d$ at first further enhances the swarming behavior, but at some point leads to its disruption. Indeed, very long delays prevent the formation of clusters. For example, this can be seen in Fig.~\ref{fig1}c (see also supplementary video 1), where $d=25$ and the clusters are less defined than in the standard Vicsek model ($d=0$, Fig.~\ref{fig1}a). A qualitative explanation of this phenomenon is the following: When $d \nu > R$ (i.e. when the agent has time to travel a distance longer than the detection radius), its reorientation occurs after it has exited the detection area; and, when $d \nu \gg R$, its reorientation is essentially uncorrelated to the orientation of the agents currently surrounding it.

We tried to use $\psi$ to quantify the effect of introducing a sensorial delay. However, as can be seen in Fig.~\ref{fig2}a, the evolution of $\psi$ for delays $d=1$ (red line) and $d=25$ (yellow line) does not show a clear difference; even the average values of $\psi$ (black dots in Fig.~\ref{fig2}b) in the steady state ($>5\,000$ timesteps) show only a trend that does not appear to be statistically significant when considering the error bars (gray area).
This is because $\psi$ measures the \emph{global} average alignment of the agents and, therefore, is reduced in the presence of multiple clusters with different directions.
In order to overcome this problem, we introduce an alternative parameter based on the Voronoi tessellation \cite{du1999centroidal}. The Voronoi tessellation partitions the plane into Voronoi cells corresponding to the agents: Each agent is associated to a Voronoi cell constituted by all those points of the plane that are closer to it than to any other agent. We define the \emph{global clustering coefficient} $c$ as the proportion of Voronoi cells whose size is smaller than a detection area:
\begin{equation} \label{eq:2}
    c = {{\rm count}\{ A_i < \pi R^2 \} \over N} \; ,
\end{equation}
where $A_i$ is the area of the $i$-th Voronoi cell. When $c \approx 0$, the systems is fully scattered, while $c \to 1$ indicates a high level of clustering. Figure~\ref{fig2}c shows the values of $c$ for delays $d=0$ (blue line), $d=1$ (red line), and $d=25$ (yellow line) as a function of the timestep: There are clear differences between their values in the steady state ($>5\,000$ timesteps).
Figure~\ref{fig2}d shows the average steady-state value (black points) and the standard deviation (gray area) of $c$ as a function of $d$. The cluster formation reaches its optimal value at $d=6$, which corresponds to the situation when agents travel a distance almost as long as a detection radius during a timestep (i.e., $d \nu \approx R$). 
As an intuitive argument, in absence of delay the detection of an agent $j$ as a neighbor of agent $i$ usually happens only when their relative distance is slightly less than $R$; instead, having a small delay allows the agent to enter deeper in the detection circle of each other, making it less likely to escape immediately after detection due to the noisy update.
For larger values of $d$, $c$ monotonically decreases because the cluster formation is prevented.
The main reason behind such phenomenon is most likely the fact that for larger delays the agents apply the gathered information after they already passed each other, thus making the information much less correlated with the current state of the system. 

\begin{figure}
\includegraphics[width=1.06\columnwidth]{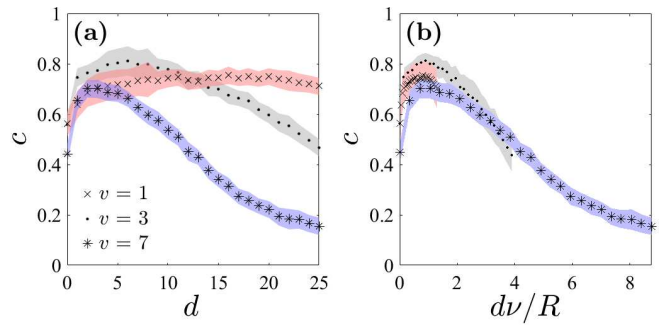}
\caption{
Dependence of clustering and swarming on the agents' speed $\nu$.
(a) The dependence of the global clustering parameter $c$ on the delay $d$ varies as a function of $\nu$ (the symbols are the averages in the steady state, and the shaded areas are the standard deviations over 10 independent runs).
The disruptive effect on the formation of the swarming clusters is more pronounced for large speeds (stars, $\nu=7$, see also supplementary video 2) than for small speeds (crosses, $\nu=1$, see also supplementary video 3).
The dots correspond to the model explored in Figs.~\ref{fig1} and \ref{fig2}. 
(b) Corresponding plot of the values of $c$ as a function of the dimensionless delay $d\nu/R$.
In all cases, we consider $N = 400$ agents with $R=20$ in a square $2\,000 \times 2\,000$ arena with periodic boundary conditions.
}
\label{fig3}
\end{figure}

The effect of the delay depends on the parameters of the system. For example, in Fig.~\ref{fig3} we explore the effect of the agent speed $\nu$, considering the cases $\nu=1$ (crosses), $3$ (dots), and $7$ (stars). In all cases, short delays enhance the clustering and swarming behavior, as alignment occurs when agents are close to each other, while long delays lead to their disruption, as alignment occurs when agents are close to the edge or completely outside of each other's detection radius. In all cases, the optimal clustering and swarming behavior occurs when $d \nu \approx R$, so that the agent has optimally entered the detection area before reorienting; this occurs at $d \approx 18$ for $\nu = 1$ (crosses), at $d \approx 6$ for $\nu = 3$ (dots), and at $d \approx 2$ for $\nu = 7$ (stars). For larger values of $d$, the reorientation of the agent occurs later when it is virtually uncorrelated to the orientation of the surrounding agents, which leads to a loss of global clustering and swarming. This can be clearly seen in Fig.~\ref{fig3}b, which show that the maximum value of $c$ is achieved for the same value of the dimensionless delay $d \nu/R$.

\begin{figure}
\includegraphics[width=1.0\columnwidth]{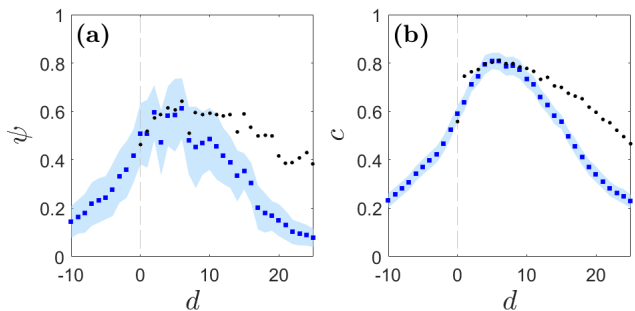}
\caption{
Negative and positive delays.
(a) Global alignment parameter $\psi$ and (b) global clustering parameter $c$ (the blue symbols are the steady-state averages and the shaded areas are the standard deviations over 10 independent runs) as a function of delay $d$. 
The black dots in (a) and (b) represent the same data as in Figs.~\ref{fig2}b and \ref{fig2}d, respectively, provided for comparison.
In all cases, we consider a Vicsek model with $N = 400$ agents ($\nu=3$, $R=20$) in a square arena ($2\,000 \times 2\,000$) with periodic boundary conditions.
See also supplementary video 4.
}
\label{fig4}
\end{figure}

It is also interesting to explore the system behavior for ``negative" delays.  A negative delay can be understood as the agents making a prediction of the future value of the orientation within their detection area in order to adjust their current orientation; importantly, this prediction can be done based on local orientations measured up to the present time. 
In practice, the agents save the orientations measured in the last five timesteps and linearly extrapolate them to the given value of the delay. When compared to the standard alignment mechanisms, this approach leads to a much smoother alignment process, as can be seen in the supplementary video 4.
Figure~4 shows the results for various values of $d$ corresponding to both positive and negative delays. The blue dots in Fig.~\ref{fig4}a show the value of $\psi$: the best alignment is achieved for $d=5$, while for larger values of $d$ and for $d<0$ the system becomes increasingly disordered.
These results are more clearly visible in the trend of $c$ (blue dots in Fig.~\ref{fig4}b): the clustering reaches its optimal value for $d=5$ and decreases for larger positive values of $d$ and for $d<0$.
For $d \ge 0$, these results are in agreement with those obtained using the measured past value of the average orientation in the detection area, as shown by the black dots in Figs.~\ref{fig4}a  and \ref{fig4}b corresponding to the values reported in Figs.~\ref{fig2}b and \ref{fig2}d, respectively. The main difference is that employing the extrapolated orientation value leads to more disordered systems. We speculate that this phenomenon is caused by the fact that the linear extrapolation leads to a wrong estimation of the orientation at a given time and therefore to a randomization of the agent orientation, as can be seen following the agents' trajectories in supplementary video 4.  

In conclusion, we have shown with numerical simulations that the introduction of a sensorial delay in the Vicsek model can alter its swarming and clustering behavior. Specifically, short positive delays lead to a more ordered and coherent motion of the ensemble of agents, while larger positive delays and negative delays lead to a disruption of the order of the system, preventing the emergence of clustering and swarming. The cluster parameter and the order parameter reach their highest values at the rescaled delay time $d \nu / R \approx 1$. Importantly, sensorial delay can potentially play a crucial role in systems different from the Vicsek model featuring alternative underlying dynamics \cite{vicsek2012collective,toner2005hydrodynamics,li2008minimal,buhl2006disorder,xu2015reversible,sanchez2012spontaneous,brambilla2013swarm,moussaid2010walking,nilsson2017metastable}. We speculate that, since some living entities, such as bacteria, are known to respond to the temporal evolution of stimuli \cite{macnab1972gradient,segall1986temporal}, the presence of a delay can already be at play in natural systems due to the time it takes to acquire, process and react to environmental information. Furthermore, we propose that engineering of sensorial delay can be employed to control and tune the clustering and swarming behavior of large ensembles of agents in applications, such as swarm robotics \cite{brambilla2013swarm},  environmental monitoring \cite{dhariwal2004bacterium,dunbabin2012robots}, and self-assembly \cite{rubenstein2014programmable,werfel2014designing}. 

GV acknowledges Anna Chiara De Luca for inspiring discussions.
This work was partially supported by the ERC Starting Grant ComplexSwimmers (grant number 677511).


%

\end{document}